\documentclass[fleqn,12pt,twoside]{article}
\usepackage{espcrc1}

\usepackage{graphicx}
\usepackage[figuresright]{rotating}

\newcommand{\AmS}{{\protect\the\textfont2
  A\kern-.1667em\lower.5ex\hbox{M}\kern-.125emS}}

\title{In medium $\sigma$ meson effects in two pion
photoproduction in nuclei.
}

\author{L. Roca\address[IFIC]{Departamento de F\'{\i}sica Te\'orica and IFIC,
Centro Mixto Universidad de Valencia-CSIC, 
Ap. Correos 22085, E-46071 Valencia, Spain}%
        \thanks{Contribution to the XVIth International Conference
	 on Particles and Nuclei (PANIC02), Osaka, Japan,
	  30 September  4 October 2002},
        E. Oset\tt\addressmark[IFIC]
        and
        M. J. Vicente Vacas\tt\addressmark[IFIC]}
       
\begin{document}

\maketitle

\begin{abstract}
We show theoretical results for $(\gamma, \pi^0 \pi^0)$ production
 on nucleons and nuclei
in the kinematical region where the scalar isoscalar $\pi \pi$ amplitude is 
influenced by the
$\sigma$ pole. The final state interaction of the pions modified by the nuclear 
medium produces a spectacular shift
of strength of the two pion invariant mass distribution induced by
 the moving of the $\sigma$ pole to lower masses and  widths as the nuclear 
 density increases.

\end{abstract}

\section{INTRODUCTION}
  
In the last years there has been an intense theoretical and
experimental debate about the nature of the $\sigma$ meson,
mostly centered on the discussion about its interpretation as an ordinary 
$q \bar{q}$ meson or a $\pi \pi $ resonance. 
The advent of $\chi PT$ showed up that the $\pi \pi$ interaction in s-wave
in the isoscalar sector is strong enough
to generate a resonance through multiple 
scattering of the pions. This seems to be the case, and even in models starting
with a seed of $q \bar{q}$ states, the incorporation of the $\pi \pi$
channels in a unitary approach leads to a large dressing by a pion
cloud which makes negligible the effects of the original $q \bar{q}$ seed.
This idea has been made more quantitative through the introduction
of the unitary extensions of $\chi PT$ ($U \chi PT$).
Even more challenging is the modification of the properties of the $\sigma$ meson at
finite nuclear density. Since present theoretical calculations agree on a sizeable modification in the
nuclear medium of the $\pi\pi$ scattering in the $\sigma$ region our purpose here 
is to find out its possible experimental signature in a very suited process
 like the $(\gamma,\pi^0 \pi^0)$ reaction in nuclei. (This contribution is a summary of
 the more extended work \cite{Roca:2002vd}).
This reaction is much better suited  than the
$(\pi,\pi\pi)$ one to investigate the
modification of the $\pi\pi$ in nuclear matter
because the photons are not distorted
by the nucleus and the reaction can test higher densities.

\section{MODEL}
For the model of the elementary $(\gamma, \pi \pi)$ reaction we follow 
 \cite{Nacher:2000eq} which considers the coupling of the photons to mesons, 
 nucleons, and the resonances
$\Delta(1232)$, $N^*(1440)$, $N^*(1520)$ and $\Delta(1700)$.
This model relies upon tree level
diagrams. Final state interaction of the $\pi N$ system is accounted for
by means of the explicit use of resonances with their widths. However,
since we do not include explicitly the $\sigma$ resonance, the final state
interaction of the two pions has to be implemented to generate it.

The $\gamma N \to N \pi^0 \pi^0$ amplitude can be decomposed in
a part which has in the final state the combination of pions
in isospin I=0 and another part where the pions are in I=2.

\begin{eqnarray}
|\pi^0_1\pi^0_2>\,=\quad
\underbrace{\frac{1}{3}|\pi^0_1\pi^0_2+\pi^+_1\pi^-_2+\pi^-_1\pi^+_2>}
_{\textrm{I=0 part}}\quad  
\underbrace{-\frac{1}{3}|\pi^0_1\pi^0_2+\pi^+_1\pi^-_2+\pi^-_1\pi^+_2>
+|\pi^0_1\pi^0_2>}_{\textrm{I=2 part}}
\label{eq:T00}
\end{eqnarray}
  
The renormalization of the 
$I=0$ $(\gamma,\pi \pi)$ amplitude is done by factorizing the on shell 
tree level $\gamma N \to \pi \pi N $ and $\pi \pi \to \pi \pi$ amplitudes in the 
loop functions. 

\begin{equation}
T_{(\gamma,\pi^0\pi^0)}^{I_{\pi\pi}=0}\to T_{(\gamma,\pi^0\pi^0)}^{I_{\pi\pi}=0}
\left(1+G_{\pi\pi}t_{\pi\pi}^{I=0}(M_I)\right)
\label{eq:GT1}
\end{equation}
where $G_{\pi\pi}$ is the loop function of the two pion propagators, 
which appears in the Bethe Salpeter equation, and $t_{\pi\pi}^{I=0}$ is the
$\pi\pi$ scattering matrix in isospin I=0, taken from \cite{Chiang:1998di}.

The multiple scattering of the two final pions can be accounted for
by means of the Bethe Salpeter equation, 

\begin{equation}
t=V+VG_{\pi\pi}t
\label{eq:Bethe}
\end{equation}
where V is given by the lowest order chiral
amplitude for $\pi\pi\to\pi\pi$ in $I=0$
and $G_{\pi\pi}$, the loop function of the two pion propagators can be
regularized by means of a cut off or with
dimensional regularization. In both approaches it has been shown
that $V$ factorizes with its on shell value in the Bethe-Salpeter equation.
 Hence, in the Bethe-Salpeter equation the integral
involving $Vt$ and the product of the two pion propagators affects only these
latter two, since $V$ and $t$ factorize outside the integral, thus leading to
Eq.~(\ref{eq:Bethe}) where $VG_{\pi\pi}t$ is the algebraic product of V, the loop 
function of the two propagators, $G_{\pi\pi}$, and the $t$ matrix.

When we renormalize the I=0 amplitude in nuclei 
to account for the pion final state
interaction, we change $G$ and $t_{\pi\pi}^{I=0}$ by their corresponding results
in nuclear matter \cite{Chiang:1998di} evaluated at the local density.
In the model of \cite{Chiang:1998di},
the $\pi\pi$ rescattering in nuclear matter was
done renormalizing the pion propagators in the medium
and introducing vertex corrections for consistency.

In the model for $( \gamma,2\pi )$ of 
\cite{Nacher:2000eq} there are indeed contact terms as implied before, as well as
other terms involving intermediate nucleon states or resonances. In this
latter case the loop function involves three propagators but 
the intermediate baryon is far off shell an the factorization of
Eq.~(\ref{eq:GT1}) still holds. There is, however, an exception in
the $\Delta$ Kroll Ruderman term, since as we increase the photon energy we get
closer to the $\Delta$ pole. For this reason this term has
been dealt separately making the
explicit calculation of the loop with one $\Delta$ and two pion propagators.

The cross section for the process in nuclei is calculated using many body techniques.
From the imaginary part of the photon selfenergy diagram with a particle-hole
excitation and two pion lines
as intermediate states, the cross section can be expressed as

\begin{eqnarray}
\sigma=&&\frac{\pi}{k}\int d^3\vec{r}\int\frac{d^{3}\vec{p}}{(2\pi)^3}
\int\frac{d^{3}\vec{q_1}}{(2\pi)^3}
\int\frac{d^{3}\vec{q_2}}{(2\pi)^3}
\, F_1(\vec{r},\vec{q_1}) F_2(\vec{r},\vec{q_2})
\frac{1}{2\omega(\vec{q_1})}\frac{1}{2\omega(\vec{q_2})}
\nonumber\\%
&&\cdot
\sum_{s_i,s_f}\overline{\sum_{pol}}\mid T\mid^{2}n(\vec{p})[1-n(\vec{k}+\vec{p}
 -\vec{q_1}-\vec{q_2})]
\nonumber\\
&&\cdot\delta(k^{0}+E(\vec{p})-\omega(\vec{q_1})-\omega(\vec{q_2})-E(\vec{k}
+\vec{p}-\vec{q_1}-\vec{q_2}))\nonumber
 \label{eq:sigma2}
\end{eqnarray}
where the factors $F_i(\vec{r},\vec{q_i})$  take into account the distortion of
the final pions in their way out trhough the nucleus and are given by

 \begin{eqnarray}\hspace{0cm}
F_i(\vec{r},\vec{q_i})=exp\left[\int_{\vec{r}}^{\infty}dl_i \frac{1}{q_i}Im
\Pi (\vec{r}_i) \right]
\label{eq:Feikonal}
\\ \nonumber & &\vspace{0.4cm}\hspace{-5.5cm}
\vec{r_i}=\vec{r}+l_i \ \vec{q_i}/\mid \vec{q_i}\mid
\end{eqnarray}

where $\Pi$ is the pion selfenergy, taken from a model based on an extrapolation for low
energy pions of a pion-nucleus optical potential developed for pionic atoms using many
body techniques. The imaginary part of the potential is split into a part that
accounts for the probability of quasielastic collisions and another one which
accounts for the pion absorption.  With this approximation
the pions which undergo absorptions are removed
from the flux but we do not remove those which undergo 
quasielastic collisions since they do not change in average
the shape or the strength of the $\pi\pi$ invariant mass distribution.

\section{RESULTS}

In the figure we can see the results for the two pion invariant mass distributions 
in the $(\gamma,\pi^0\pi^0)$ and $(\gamma,\pi^{\pm}\pi^0)$ reactions
 on $^1H$, $^{12}C$ and $^{208}Pb$.  The difference between the 
solid and dashed curves is the use of the in
medium $\pi \pi$ scattering and $G$ function instead of the free ones, which
we take from \cite{Chiang:1998di}.  As one can see in the figure, there 
is an appreciable shift of strength to the low
invariant mass region due to the in medium  $\pi \pi$  interaction.   This
shift is remarkably similar to the one found in the preliminary measurements of
\cite{Messchendorp:2002au}. This shift is not seen in the $(\gamma,\pi^{\pm}\pi^0)$ channel
 because the $\pi^{\pm}\pi^0$ are not allowed to be in isospin I=0.

These results show a clear signature of the modified $\pi\pi$
interaction in the medium.  The fact that the photons are not distorted has
certainly an advantage over the pion induced reactions and allows one to see 
inner parts of the nucleus.

Although we have been discussing the $\pi \pi$ interaction in the nuclear
medium it is clear that we can relate it to the modification of the $\sigma$
in the medium. We have mentioned that the reason for the shift of strength to
lower invariant masses in the mass distribution is due to the accumulated
strength in the scalar isocalar $\pi \pi$ amplitude in the medium. Yet, this
strength is mostly governed by the presence of the $\sigma$ pole and there have
been works suggesting that the sigma should move to smaller masses and widths
when embedded in the nucleus.
 The present results
represent an evidence of the move of pole position of the $\sigma$ moves
to smaller energies as the nuclear density increases, a
phenomenon which would come to strengthen once more the nature of the $\sigma$
meson as dynamically generated by the multiple scattering of the pions through
the underlying chiral dynamics.

\newpage

\begin{figure}
\center\includegraphics[angle=0,width=11cm]{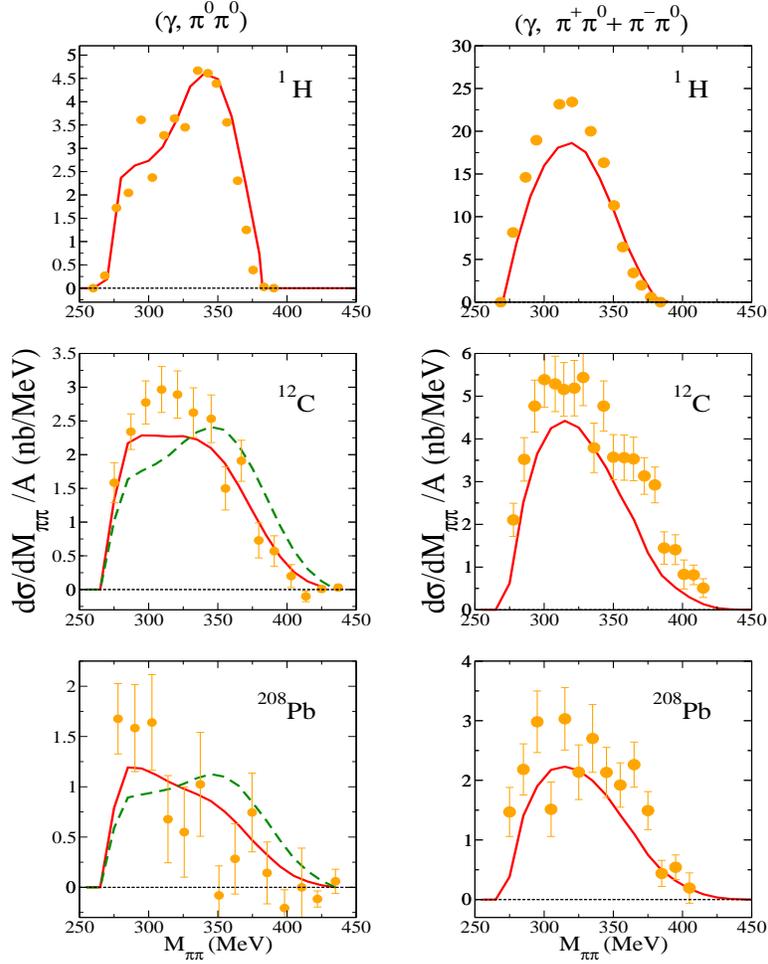}
\caption{\small{Two pion invariant mass distribution for $\pi^0\pi^0$ and $\pi^{\pm}\pi^0$
 photoproduction 
in $^{12}C$ and $^{208}Pb$.
Continuous lines: using the in medium final $\pi\pi$ interaction. 
Dashed lines: using the final $\pi\pi$ interaction at zero
density.
Exp. data from \cite{Messchendorp:2002au}}}
\end{figure}


\begin{thebibliography}{99}

\bibitem{Roca:2002vd}
L.~Roca, E.~Oset and M.~J.~Vicente Vacas,
Phys.\ Lett.\ B {\bf 541} (2002) 77
[arXiv:nucl-th/0201054].

\bibitem{Nacher:2000eq}
J.~C.~Nacher, E.~Oset, M.~J.~Vicente and L.~Roca,
Nucl.\ Phys.\ A {\bf 695} (2001) 295
[arXiv:nucl-th/0012065].

\bibitem{Chiang:1998di}
H.~C.~Chiang, E.~Oset and M.~J.~Vicente-Vacas,
Nucl.\ Phys.\ A {\bf 644} (1998) 77
[arXiv:nucl-th/9712047].

\bibitem{Messchendorp:2002au}
J.~G.~Messchendorp {\it et al.},
Phys.\ Rev.\ Lett.\  {\bf 89} (2002) 222302
[arXiv:nucl-ex/0205009].

\end{thebibliography}
\end{document}